\documentclass[utf8]{frontiersSCNS} 
\usepackage{epstopdf}
\usepackage{comment}
\usepackage{url,hyperref,lineno,microtype,subcaption}
\usepackage[onehalfspacing]{setspace}

\def\keyFont{\fontsize{8}{11}\helveticabold }
\def\firstAuthorLast{Qiaoyu Tan {et~al.}} 
\def\Authors{Qiaoyu Tan\,$^{}$, Ninghao Liu\,$^{}$ and Xia Hu\,$^{*}$}
\def\Address{Department of Computer Science and Engineering, Texas A\&M University
}
\def\corrAuthor{Corresponding Author}

\def\corrEmail{\{qytan, nhliu43, xiahu\}@tamu.edu}

\begin{document}
\onecolumn
\firstpage{1}

\title[Deep Representation Learning for Social Network Analysis]{Deep Representation Learning for Social Network Analysis} 

\author[\firstAuthorLast ]{\Authors} 
\address{} 
\correspondance{} 

\extraAuth{}

\maketitle
\begin{abstract}
Social network analysis is an important problem in data mining. A fundamental step for analyzing social networks is to encode network data into low-dimensional representations, i.e., network embeddings, so that the network topology structure and other attribute information can be effectively preserved. Network representation leaning facilitates further applications such as classification, link prediction, anomaly detection and clustering. In addition, techniques based on deep neural networks have attracted great interests over the past a few years. In this survey, we conduct a comprehensive review of current literature in network representation learning utilizing neural network models. First, we introduce the basic models for learning node representations in homogeneous networks. Meanwhile, we will also introduce some extensions of the base models in tackling more complex scenarios, such as analyzing attributed networks, heterogeneous networks and dynamic networks. Then, we introduce the techniques for embedding subgraphs. After that, we present the applications of network representation learning. At the end, we discuss some promising research directions for future work. 

\tiny
 \keyFont{ \section{Keywords:} Deep learning, social networks, deep social network analysis, representation learning, network embedding} 
\end{abstract}

\section{Introduction}
Social networks, such as Facebook, Twitter and Linkedin, have greatly facilitated communications between web users around the world. The analysis of social networks helps summarize the interests and opinions of users (nodes), discovering patterns from the interactions (links) between users, and mining the events that take place in online platforms. The information obtained by analyzing social networks could be especially valuable for many applications. Some typical examples include online advertisement targeting~\citep{li2015real}, personalized recommendation \citep{song2006personalized}, viral marketing~\citep{chen2010scalable, leskovec2007dynamics}, social healthcare~\citep{tang2012ranking}, social influence analysis~\citep{peng2017social}, academic networks analysis~\citep{dietz2007unsupervised, guo2014two}.

One central problem in social network analysis is how to extract useful features from non-Euclidean structured networks, to enable the deployment of downstream machine learning prediction models for specific analysis. For example, in the case of recommending new friends to a user in a social network, the key challenge might be how to embed network users into a low-dimensional space so that the closeness between users could be easily measured with distance metrics. To process the structure information in networks, most previous efforts mainly rely on hand-crafted features, such as kernel functions~\citep{vishwanathan2010graph}, graph statistics (i.e., degrees or clustering coefficients)~\citep{bhagat2011node}, or other carefully engineered features~\citep{liben2007link}. However, such feature engineering process could be very time-consuming and expensive, making it ineffective for many real-world applications. An alternative way to avoid this limitation is to automatically learn feature representations that capture various information sources in networks~\citep{bengio2013representation,liao2018attributed}. 
The goal is to learn a transformation function that maps nodes, subgraphs or even the whole network as vectors to a low-dimensional feature space, where the spatial relations between the vectors reflect the structures or contents in the original network. Given these feature vectors, subsequent machine learning models such as classification models, clustering models and outlier detection models could be directly used towards target applications.

Along with the substantial performance improvement gained by deep learning on image recognition, text mining, and natural language processing tasks~\citep{bengio2009learning}, developing network representation methods using neural network models have received increasing attentions in recent years. 
In this survey, we provide a comprehensive overview of recent advancements in network representation learning using neural network models. After introducing the notations and problem definitions, we first review the basic representation learning models for node embedding in homogeneous networks. Specifically, based on the type of representation generation modules, we divide the existing approaches into three categories: embedding look-up based, autoencoder based and graph convolution based. Then, we give an overview of approaches that learn representations for subgraphs in networks, which to some extent rely on the techniques of node representation learning. After that, we list some applications of network representation models. At the end, we discuss some promising research directions for future work. 
 
\section{Notations and Problem Definitions}
In this section, we define some important terminologies that will be used in later sections, and then give the formal definition of network representation learning problem. In general, we use boldface uppercase letters (e.g., $\mathbf{A}$) to denote matrices, boldface lowercase letters (e.g., $\mathbf{a}$) to denote vectors, and lowercase letters (e.g., $a$) to denote scalars. The $(i,j)$ entry, the $i$-th row and the $j$-th column of a matrix $\mathbf{A}$ is denoted as $\mathbf{A}_{ij}$, $\mathbf{A}_{i*}$ and $\mathbf{A}_{*j}$, respectively.

\textit{Definition 1 (Network)}. Let $\mathcal{G}=\{ \mathcal{V}, \mathcal{E}, \mathbf{X}, \mathbf{Y} \}$ be a network, where the $i$-th node (or vertex) is denoted as $v_i \in \mathcal{V}$ and $e_{i,j} \in \mathcal{E}$ denotes the edge between node $v_i$ and $v_j$. $\mathbf{X}$ and $\mathbf{Y}$ are node attributes and labels, if available. Besides, we let $\mathbf{A}\in\mathbb{R}^{N\times{N}}$ denote the associated adjacency matrix of $\mathcal{G}$. $\mathbf{A}_{ij}$ is the weight of $e_{i,j}$, where $\mathbf{A}_{ij}> 0$ indicates that the two nodes are connected, and otherwise $\mathbf{A}_{ij}=0$. For undirected graphs, $\mathbf{A}_{ij}=\mathbf{A}_{ji}$.

In many scenarios, the nodes and edges in $\mathcal{G}$ can also be associated with type information. Let $\tau_v: \mathcal{V} \rightarrow T^v$ be a node-type mapping function and $\tau_e: \mathcal{E} \rightarrow T^e$ be an edge-type mapping function, where $T^v$ and $T^e$ denote the set of node and edge types, respectively. Here, each node $v_i\in \mathcal{V}$ has one specific type, e.g., $\tau_v(v_i)\in T^v$. Similarly, for each edge $e_{ij}$, $\tau_{e}(e_{ij})\in T^e$.

\textit{Definition 2 (Homogeneous Network)}. A homogeneous network is a network in which $|T^v|=|T^e|=1$. All nodes and edges in $G$ belong to one single type. 

\textit{Definition 3 (Heterogeneous Network)}. A heterogeneous network is a network with $|T^v|+|T^e|>2$.  
There are at least two different types of nodes or edges in heterogeneous networks. 

Given a network $\mathcal{G}$, the task of network representation learning is to train a mapping function $f$ that maps certain components in $\mathcal{G}$, such as nodes or subgraphs, into a latent space. Let $D$ be the dimension of the latent space and usually $D \ll |\mathcal{V}|$. In this work, we focus on the problem of node representation learning and subgraph representation learning.

\textit{Definition 4 (Node Representation Learning)}. Suppose $\textbf{z}\in \mathbb{R}^D$ denote the latent vector of node $v$, node representation learning aims to build a mapping function $f$ so that $\textbf{z} = f(v)$. It is expected that nodes with similar roles or characteristics, which is defined according to specific application domains, are mapped close to each other in the latent space.

\textit{Definition 5 (Subgraph Representation Learning)}. Let $g$ denote a subgraph of $\mathcal{G}$. The nodes and edges in $g$ are denoted as $\mathcal{V}_S$ and $\mathcal{E}_S$, respectively, and we have $\mathcal{V}_S \subset \mathcal{V}$ and $\mathcal{E}_S \subset \mathcal{E}$. The subgraph representation learning aims to learn a mapping function $f$ so that $\textbf{z} = f(g)$, where in this case $\textbf{z}\in \mathbb{R}^D$ corresponds to the latent vector of $g$.

\begin{figure}[h!]
\centering
\includegraphics[width=16cm]{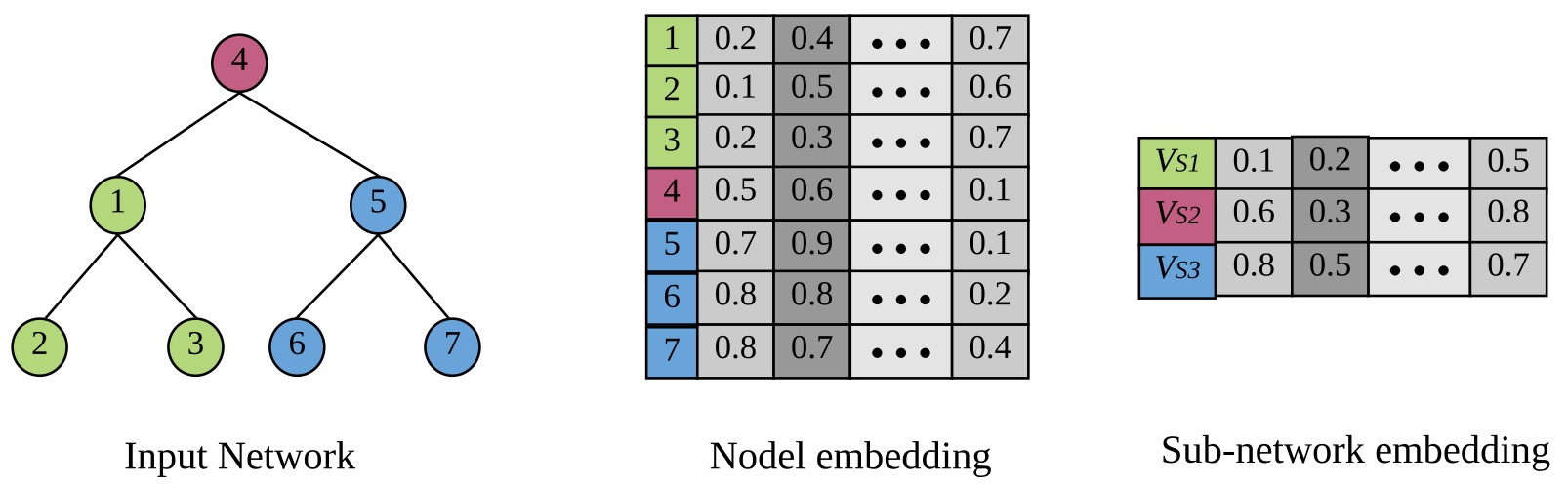}
\caption{A toy example of node representation learning and subgraph representation learning (best viewed in color). There are three subgraphs in the input network denoted by different colors. The target of node embedding is to generate one representation for each individual node, while subgraph embedding is to learn one representation for an entire subgraph.}\label{fig1}
\end{figure}

Figure~\ref{fig1} shows a toy example of network embedding. There are three subgraphs in this network distinguished with different colors: $\mathcal{V}_{S_1}=\{v_1,v_2,v_3\}$, $\mathcal{V}_{S_2}=\{v_4\}$, and $\mathcal{V}_{S_3}=\{v_5,v_6,v_7\}$. Given a network as input, the example below generates one representation for each node, as well as for each of the three subgraphs.

\section{Neural Network Based Models}
\label{section3}
Neural networks have been demonstrated to have powerful capabilities in capturing complex patterns in data, and have achieved substantial success in the fields of computer vision, audio recognition and natural language processing, etc. Recently, some efforts have been made to extend neural network models to learn representations from network data. Based on the type of base neural networks that are applied, we categorize them into three subgroups: look-up table based models, autoencoder based models, and GCN based models.
In this section, we first give an overview of network representation learning from the perspective of \textit{encoding} and \textit{decoding}. Then we discuss the details of some well-known network embedding models and how they fulfill the two steps.  In this section, we only discuss representation learning for nodes. The models dealing with subgraphs will be introduced in later sections.

\subsection{Framework Overview from the Encoder-Decoder Perspective}
In order to elaborate the diversity of various neural network architectures, we argue that different techniques can be derived from the aspect of \textit{encoding} and \textit{decoding} schema, as well as their \textit{target network structure} constrained for low dimensional feature space. 
Specifically, existing methods can be reduced to solving the following optimization problem:
\begin{equation}
\min \limits_{\Psi} \sum_{\phi \in \Phi_{tar}} \mathcal{L} (\psi_{dec}(\psi_{enc}(\mathcal{V}_{\phi})), \phi | \Psi),
\label{eq1}
\end{equation}
where $\Phi_{tar}$ is the target relations that the embedding algorithm expects to preserve, and $\mathcal{V}_{\phi}$ denotes the nodes involved in $\phi$. $\psi_{enc}: \mathcal{V} \rightarrow \mathcal{R}^{D}$ is the \textit{encoding} function that maps nodes into representation vectors, and $\psi_{dec}$ is a decoding function that reconstructs the original network structure from the representation space. $\Psi$ denotes the trainable parameters in encoders and decoders.
By minimizing the loss function above, model parameters are trained so that the desired network structure $\Psi_{tar}$ are preserved. As we will show in subsequent sections, from the overview framework aspect, the primary distinctions between various network representation methods rely on how they define the three components. 

\begin{figure}[t]
\begin{center}
\includegraphics[width=16cm]{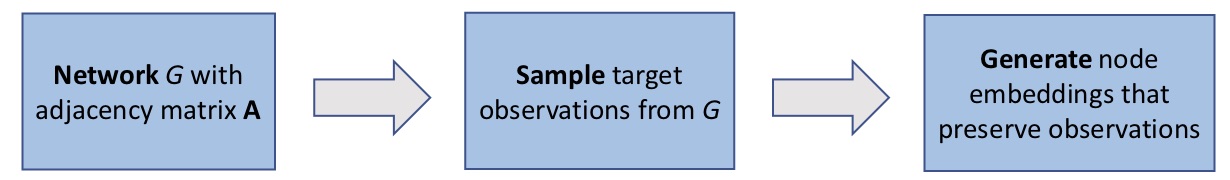}
\end{center}
\caption{Building blocks of models with embedding look-up tables. There are two key components of these work: \textit{sampling} and \textit{generating}. The primary distinctions between different methods under this line rely on how to define the two components. }
\label{fig2}
\end{figure}

\subsection{Models with Embedding Look-up Tables}
Instead of using multiple layers of nonlinear transformation, network representation learning could be achieved simply using look-up tables which directly map a node index into its corresponding representation vector. Specifically, a look-up table could be implemented using a matrix, where each row corresponds to the representation of one node. The diversity of different models mainly lies in the definition of target relations in the network data that we hope to preserve. In the rest of this subsection, we will first introduce DeepWalk~\citep{perozzi2014deepwalk} to discuss the basic concepts and techniques in network embedding, and then extend the discussion to more complex and practical scenarios.

\textbf{Skip-Gram Based Models}. As a pioneering network representation model, DeepWalk treats nodes as words, samples random walks as sentences, and utilizes the skip-gram model~\citep{mikolov2013efficient} to learn the representations of nodes as shown in Figure~\ref{fig2}. In this case,
the encoder $\psi_{enc}$ is implemented as two embedding look-up tables $\mathbf{Z}\in \mathbb{R}^{N\times D}$ and $\mathbf{Z}^c\in \mathbb{R}^{N\times D}$, respectively for target embeddings and context embeddings. The network information $\phi\in \Phi_{tar}$ that we try to preserve is defined as the node-context pairs $(v_i, \mathcal{N}(v_i))$ observed in the random walks, where $\mathcal{N}(v_i)$ denotes the context nodes (or neighborhood) of $v_i$. The objective is to maximize the probability of observing a node's neighborhood conditioned on embeddings:
\begin{equation}
\mathcal{L} = - \sum_{v_i\in \mathcal{V}}\sum_{v_j \in \mathcal{N}(v_i)} \log p( \textbf{e}_j \mathbf{Z}^c | \textbf{e}_i \mathbf{Z} ), 
\label{eq2}
\end{equation}
where $\textbf{e}_i $ is a one-hot row vector of length $N$ that picks the $i$-th row of $\mathbf{Z}$. 
Let $\mathbf{z}_i = \textbf{e}_i \mathbf{Z} $ and $\mathbf{z}^c_j = \textbf{e}_j \mathbf{Z}^c $, the conditional probability above is formulated as
\begin{equation}
p( \mathbf{z}^c_j | \mathbf{z}_i) =\frac{\exp({\mathbf{z}^c_{j}}\mathbf{z}^T_i)}{\sum_{k=1}^{|\mathcal{V}|}\exp({\mathbf{z}^c_{k}}\mathbf{z}^T_i)},
\label{eq3}
\end{equation}
so that $\psi_{dec}$ could be regarded as link reconstruction based on the normalized proximity between different nodes. In practice, the computation of the probability is expensive due to the summation over every node in the network, but hierarchical softmax or negative sampling can be applied to reduce time complexity.

There are also some approaches that are developed based on similar ideas. LINE~\citep{tang2015line} defines the first-order and second-order proximity for learning node embedding, where the latter can be seen as a special case of DeepWalk with context window length set as $1$. Meanwhile, node2vec~\citep{grover2016node2vec} applies different random walk strategies, which provides a trade-off between breadth-first search (BFS) and depth-first search (DFS) in networks search strategies. Planetoid~\citep{yang2016revisiting} extends skip-gram models for semi-supervised learning, which predicts the class label of nodes along with the context in the input network data. In addition, it has been shown that there exists a close relationship between skip-gram models and matrix factorization algorithms~\citep{qiu2018network, levy2014neural}. Therefore, network embedding models that utilize matrix factorization techniques, such as LE~\citep{belkin2002laplacian}, Grarep~\citep{cao2015grarep}, and HOPE~\citep{ou2016asymmetric}, may also be implemented in the similar manner. Random sampling based approaches have the capacity to allow a flexible and stochastic measure of node similarity, making them not only achieve higher performance in many applications but also become more scalable toward large-scale datasets.

\textbf{Attributed Network Embedding Models.} Social networks are rich in side information, where nodes could be associated with various attributes that characterize their properties. Inspired by the idea of inductive matrix completion~\citep{natarajan2014inductive}, TADW~\citep{yang2015network} extends the framework of DeepWalk by incorporating features of vertices into network representation learning. Besides sampling from plain networks, FeatWalk~\citep{huang2019large} proposes a novel feature-based random walk strategy to generate node sequences by considering node similarity on attributes. With the random walks based on both topological and attribute information, the skip-gram model is then applied to learn node representations. 

\textbf{Heterogeneous Network Embedding Models.} Nodes in networks could be of different types, which poses the challenge of how to preserve relations among them. HERec~\citep{shi2019heterogeneous} and metapath2vec++~\citep{dong2017metapath2vec} propose meta-path based random walk schema to discover the context across different types of nodes.
The skip-gram architecture in metapath2vec++ is also modified, so that the normalization term in softmax only consider the nodes of the same type as the target node. In a more complex scenario where we have both nodes and attributes of different types, HNE~\citep{chang2015heterogeneous} combines feed-forward neural networks and embedding models towards a unified framework. Suppose $\mathbf{z}^a$ and $\mathbf{z}^b$ denote the latent vectors of two different types of nodes, HNE defines two additional transformation matrices $\mathbf{U}$ and $\mathbf{V}$ to respectively map $\mathbf{z}^a$ and $\mathbf{z}^b$ to the joint space. Let $v_i, v_j \in \mathcal{V}_a$ and $v_k, v_l \in \mathcal{V}_b$, intra-type node similarity and inter-type node similarity are defined as
\begin{equation}
s(v_i, v_j) = \mathbf{z}^a_i \mathbf{U} (\mathbf{z}^a_j \mathbf{U})^T, \,\,\,\,\,\,
s(v_i, v_k) = \mathbf{z}^a_i \mathbf{U} (\mathbf{z}^b_k \mathbf{V})^T, \,\,\,\,\,\,
s(v_k, v_l) = \mathbf{z}^b_k \mathbf{V} (\mathbf{z}^b_l \mathbf{V})^T,
\end{equation}
where we hope to preserve various types of similarities during training. As for obtaining $\mathbf{z}^a$ and $\mathbf{z}^b$, HNE applies different feed-forward neural networks to map raw input (e.g., images and texts) to latent spaces, thus enables an end-to-end training framework. Specifically, the authors use a CNN to process images and a fully-connected neural network to process texts.

\textbf{Dynamic Embedding Models.} Real world social networks are not static and will evolve over time with addition/deletion of nodes and links. To deal with this challenge, DNE~\citep{du2018dynamic} presents a decomposable objective to learn the representation of each node separately, where the impact of network changes on existing nodes is measurable and the greatly affected nodes will be chosen for update as learning process proceeds. In addition, DANE~\citep{Li2017Attributed} leverages matrix perturbation theory for tackling online embedding updates.

\begin{figure}[t]
\begin{center}
\includegraphics[width=16cm]{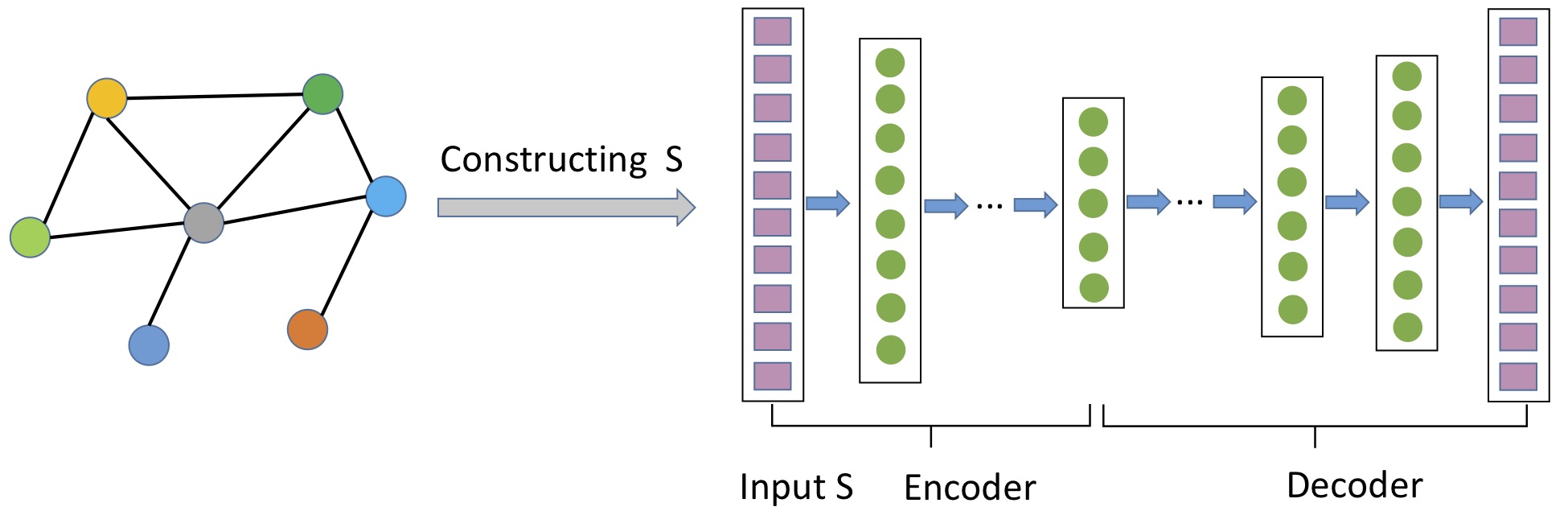}
\end{center}
\caption{An example of autoencoder-based network representation algorithms. Rows of the proximity matrix $\mathbf{S}\in \mathbb{R}^{|\mathcal{V}|\times{|\mathcal{V}|}}$ are fed into the autoencoder to learn and generate embeddings $\mathbf{Z}\in \mathbb{R}^{|\mathcal{V}|\times{D}}$ at the hidden layer. }
\label{fig3}
\end{figure}

\subsection{Autoencoder Techniques}
In this section, we discuss network representation models based on the autoencoder architecture~\citep{hinton2006reducing,bengio2013representation}. As shown in Figure~\ref{fig3}, an autoencoder consists of two neural network modules: encoder and decoder. The encoder $\psi_{enc}$ maps the features of each node into a latent space, and the decoder $\psi_{doc}$ reconstructs the information about the network from the latent space. Usually the hidden representation layer has a smaller size than that of the input/output layer, forcing it to create a compressed representation that captures the non-linear structure of network. Formally, following Equation~\ref{eq1}, the objective function of autoencoder is to minimize the reconstruction error between the input and the output decoded from low-dimensional representations. 

\textbf{Deep Neural Graph Representation (DNGR)}. DNGR~\citep{cao2016deep} attempts to preserve a node's local neighborhood information using a stacked denoising autoencoder. Specifically, assume $\mathbf{S}$ is the PPMI matrix~\citep{bullinaria2007extracting} constructed from $\mathbf{A}$, then DNGR minimizes the following loss: 
\begin{equation}
\mathcal{L}=\sum_{v_i\in \mathcal{V}}||\psi_{dec}(\mathbf{z}_i) - \mathbf{S}_{i*}||^2_2 \ \ \ s.t. \ \ \mathbf{z}_i = \psi_{enc}(\mathbf{S}_{i*}),
\label{eq5}
\end{equation}
where $\mathbf{S}_{i*} \in \mathbb{R}^{|\mathcal{V}|}$ denotes the associated neighborhood information of $v_i$. In this case, $\Phi_{tar} = \{ \mathbf{S}_{i*} \}_{v_i\in V}$ and DNSR targets to reconstruct the PPMI matrix. 
$\mathbf{z}_i$ is the embedding of node $v_i$ in hidden layer. 

\textbf{Structural Deep Network Embedding (SDNE)}. SDNE~\citep{wang2016structural} is another autoencoder-based model for network representation learning. The objective function of SDNE is:
\begin{equation}
\mathcal{L}=\sum_{v_i \in V}|| (\psi_{dec}(\mathbf{z}_i)-\mathbf{S}_{i*})\odot \mathbf{b}_i ||^2_2+\sum_{i, j=1}^{|V|}\mathbf{S}_{ij}||\mathbf{z}_{i}-\mathbf{z}_{j}||^2_2, \ \ \Psi_{tar}=\{\mathbf{S}_{i*}, \mathbf{S}_{ij}\} .
\label{eq6}
\end{equation}
The first term is an autoencoder as in Equation~\ref{eq5}, except that the recostruction error is weighted, so that more emphasis is put on recovering non-zero entries in $\mathbf{S}_{i*}$. The second part is motivated by Laplacian Eigenmaps that imposes nearby nodes to have similar embeddings. Besides, SDNE differs from DNGR in the definition of $\mathbf{S}$, where DNGR defines $\mathbf{S}$ as the PPMI matrix while SDNE sets $\mathbf{S}$ as the adjacency matrix.

It is worth noting that, unlike in Equation~\ref{eq2} that uses one-hot indicator vector for embedding look-up, DNGR and SDNE transform each node's information to an embedding by training neural network modules. Such distinction allows autoencoder-based methods to directly model on a node's neighborhood structure and features, which is not straightforward for random walk approaches. Therefore, it is straightforward to incorporate richer information sources (e.g., node attributes) into representation learning, as to be introduced below. However, autoencoder-based methods may suffer from scalability issues as the input dimension is $|\mathcal{V}|$, which may result in significant time costs in real massive datasets.

\textbf{Autoencoder-Based Attributed Network Embedding.} The structure of autoencoders facilitates the incorporation of multiple information sources towards joint representation learning. Instead of only mapping nodes to the latent space, CAN~\citep{meng2019co} proposes to learn the representation of nodes and attributes in the same latent space by using variational autoencoders (VAEs)~\citep{doersch2016tutorial}, in order to capture the affinities between nodes and attributes. DANE~\citep{gao2018deep} utilizes the correlation between topological and attribute information of nodes by building two autoencoders for each information source, and then encourages the two sets of latent representations to be consistent and complementary. ~\citep{li2017variation} adopts another strategy, where topological feature vector and content information vector (learned by doc2vec~\citep{le2014distributed}) are directly concatenated and put into a VAE to capture the nonlinear relationship between them.

\begin{figure}[t]
\begin{center}
\includegraphics[width=14cm,height=6.5cm]{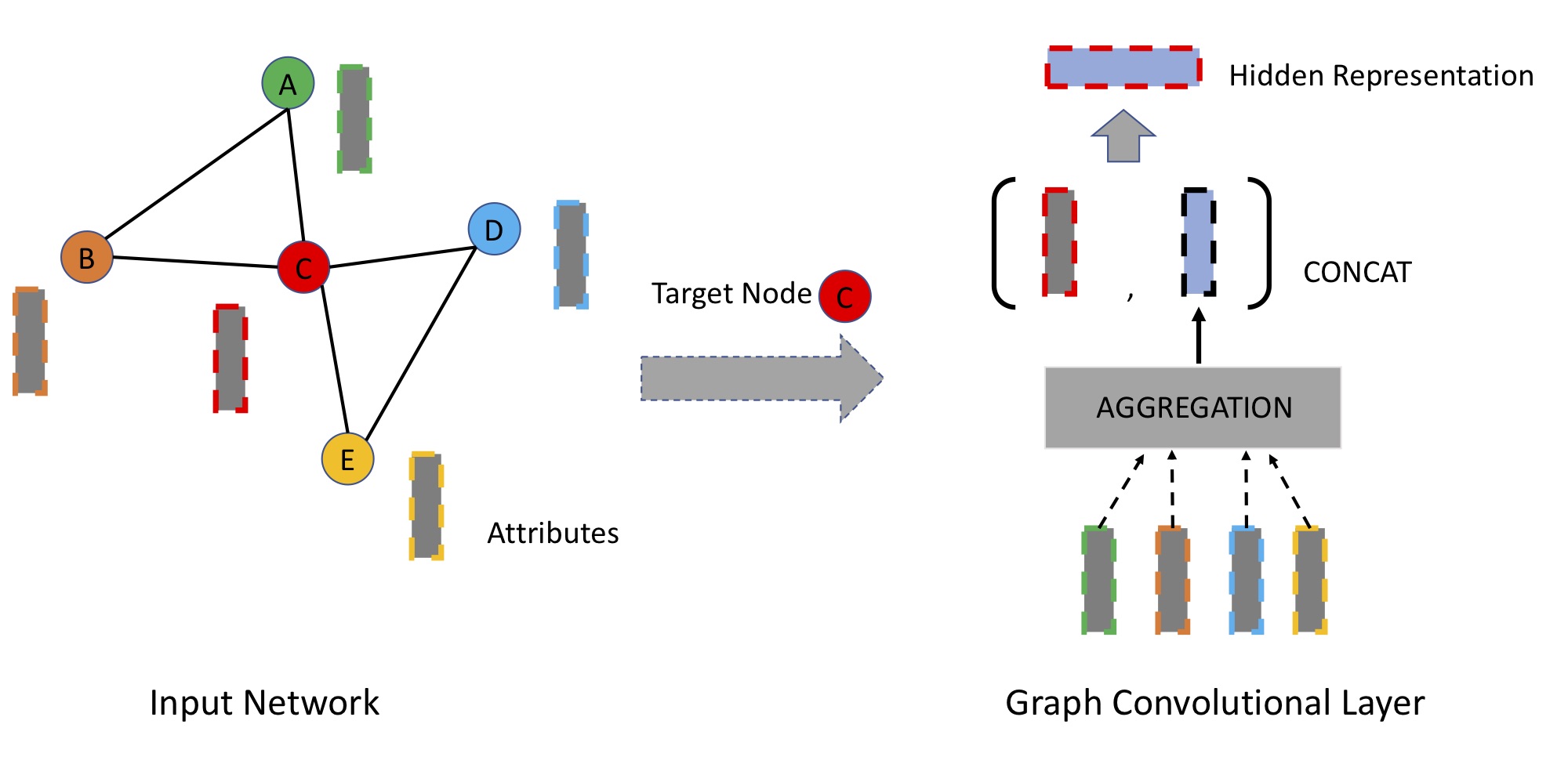}
\end{center}
\caption{An overview of graph convolutional networks. The dashed rectangles denote node attributes. The representation of each individual node (e.g., node C) is aggregated from its immediate neighbors (e.g., node A, B, D, E), concatenated with the lower-layer representation of itself.}
\label{fig4}
\end{figure}

\subsection{Graph Convolutional Approaches}
Inspired by the significant performance improvement of convolutional neural networks (CNN) in image recognition, recent years have witnessed a surge in adapting convolutional modules to learn representations of network data. The intuition behind is to generate node embedding by aggregating information from its local neighborhood as shown in Figure~\ref{fig4}. Different from autoencoder-based approaches, the encoding function of graph convolutional approaches leverages a node's local neighborhood as well as attribute information. Some efforts~\citep{bruna2013spectral,henaff2015deep,defferrard2016convolutional,hamilton2017inductive} have been made to extend traditional convolutional networks for network data to generate network embedding in the past few years. The convolutional filters of these approaches are either spatial filters or spectral filters. Spatial filters operate directly on the adjacency matrix whereas spectral filters operate on the spectrum of graph Laplacian~\citep{defferrard2016convolutional}.

\textbf{Graph Convolutional Networks (GCN)}. GCN~\citep{bronstein2017geometric} is a well-known semi-supervised graph convolutional networks. It defines a convolutional operator on network, and iteratively aggregates embeddings of neighbors of a node and uses the aggregated embedding as well as its own embedding at previous iteration to generate the node's new representation. The layer-wise propagation rule of encoding function $\psi_{enc}$ is defined as:
\begin{equation}
\mathbf{H}^{k}=\sigma(\hat{\mathbf{D}}^{-\frac{1}{2}}\hat{\mathbf{A}}\hat{\mathbf{D}}^{-\frac{1}{2}}\mathbf{H}^{k-1}\mathbf{W}^{k-1}),
\label{eq7}
\end{equation}
where $\mathbf{H}^{k-1}$ denotes the learned embeddings in layer $k-1$, and $\mathbf{H}^0=\mathbf{X}$. $\hat{\mathbf{A}}=(\mathbf{I}_G + \mathbf{A})$ is the adjacency matrix with added self-connections. $\mathbf{I}_G$ is the identity matrix, $\hat{\mathbf{D}}_{ii}=\sum_{j}\hat{\mathbf{A}}_{ij}$. $\mathbf{W}^{k-1}$ is a layer-wise trainable weight matrix. $\sigma(\cdot)$ denotes an activation function such as ReLU. The loss function for supervised training is to evaluate the cross-entroy error over all labeled nodes:
\begin{equation}
\mathcal{L}=-\sum_{v_i\in \mathcal{V}}\sum_{f=1}^F\mathbf{Y}_{if}\ln \hat{\mathbf{Y}}_{if}, \ \ s.t. \ \ \hat{\mathbf{Y}}=\psi_{dec}(\mathbf{Z}), \ \ \mathbf{Z}=\psi_{enc}(\mathbf{X}, \mathbf{A}), 
\label{eq8}
\end{equation}
where $\hat{\mathbf{Y}}\in\mathbb{R}^{N\times{F}}$ is the predictive matrix with $F$ candidate labels. $\psi_{dec}(\cdot)$ can be viewed as a fully-connected network with the softmax activation function to map representations to predicted labels. 
Note that unlike autoencoders that explicitly treat each node's neighborhood as features or reconstruction goals as in Equation~\ref{eq5} or Equation~\ref{eq6}, GCN implicitly applies the local neighborhood links on each encoding layer as pathways to aggregate embeddings from neighbors, so that higher order network structures are utilized. Since Equation~\ref{eq8} is a supervised loss function, $\Phi_{tar}$ is not applicable here. However, the loss function can also be formulated in unsupervised manners, similar to the skip-gram model~\citep{hamilton2017inductive,kipf2016variational}. GCN may suffer from the scalability problem when the size of $\mathbf{A}$ is large. The corresponding training algorithms have been proposed to tackle this challenge~\citep{ying2018graph}, where the network data is processed in small batches and we can sample a node's local neighbors instead of using all of them.

\textbf{Inductive Training With GCN.} So far many basic models we have reviewed mainly generate network representations in a transductive manner. GraphSAGE~\citep{hamilton2017inductive} emphasized the inductive capability of GCN. Inductive learning is essential for high-throughput machine learning systems, especially when operating on evolving networks that constantly encounter unseen nodes~\citep{yang2016revisiting,guo2018spine}.
The core representation update scheme of GraphSAGE is similar to that of traditional GCN, except that the operation on the whole network is replace by sample-based representation aggregators:
\begin{equation}
\mathbf{h}^{k}_i=\sigma(\mathbf{W}^k \cdot \textsc{CONCAT}(\mathbf{h}^{k-1}_i, \textsc{AGGREGATE}_k(\{\mathbf{h}^{k-1}_j,  \forall j\in \mathcal{N}(v_i)\}))) ,
\label{eq9}
\end{equation}
where $\mathbf{h}_i^k$ is the hidden representation of node $v_i$ in the $k$-th layer. $\textsc{CONCAT}$ denotes concatenation operator and $\textsc{AGGREGATE}_k$ represents neighborhood aggregation function of the $k$-th layer (e.g., element-wise mean or max operator). $\mathcal{N}(v_i)$ denotes the neighbors of $v_i$. Compared with Equation~\ref{eq7}, GraphSAGE only needs to aggregate feature vectors from the partial set of neighbors, making it scalable for large-scale data.
Given the attribute features and neighborhood relations of an unseen node, GraphSAGE can generate the embedding of this node by leveraging its local neighbors as well as attributes via forward propagation. 

\textbf{Graph Attention Mechanisms.} Attention mechanisms have become the standard technique in many sequence-based tasks, in order to make models focus on the most relevant parts of the input in making decisions. We could also utilize attention mechanisms to aggregate the most important features from nodes' local neighbors.
GAT~\citep{velickovic2017graph} extends the framework of GCN by replacing the standard aggregation function with an attention layer to aggregate message from most important neighbors. Also, ~\citep{thekumparampil2018attention} proposes to remove all intermediate fully-connected layers in conventional GCN, and replace the propagation layers with attention layers. It thus allows the model to learn a dynamic and adaptive local summary of neighborhoods, greatly reduces the parameters, and also achieves more accurate predictions.

\section{Subgraph Embedding}
Besides learning representations for nodes, recent years have also witnessed an increasing branch of research efforts that try to learn representations for a set of nodes and edges as an integral. Thus, the goal is to represent a subgraph with a low-dimensional vector. Many traditional methods that operate on subgraphs rely on graph kernels~\citep{haussler1999convolution}, which decompose a network into some atomic substructures such as graphlets, subtree patterns and paths, and treat these substructures as features to obtain an embedding through further transformation. 
In this section, however, we focus on reviewing methods that seek to automatically learn embeddings of subgraphs using deep models. For those who are interested in graph kernels, we refer the readers to~\citep{vishwanathan2010graph}. 

According to the literature, most existing methods are built upon the techniques used for node embedding, as introduced in Section~\ref{section3}. However, in graph representation problems, the label information is associated with particular subgraphs instead of individual nodes or links. In this survey, we divide the approaches of subgraph representation learning into two categories based on how they aggregate node-level embeddings in each subgraph. The detailed discussion for each category is as below.

\subsection{Flat Aggregation}
Assume $\mathcal{V}_{\mathcal{S}}$ denotes the set of nodes in a particular subgraph and $\mathbf{z}_{\mathcal{S}}$ represents the subgraph's embedding, $\mathbf{z}_{\mathcal{S}}$ could be obtained by aggregating the embeddings of all individual nodes in the subgraph: 
\begin{equation}
\mathbf{z}_{\mathcal{S}}=\psi_{aggr}(\{\mathbf{z}_i, v_i \in \mathcal{V}_{\mathcal{S}}\}),
\label{eq11}
\end{equation}
where $\psi_{aggr}$ denotes the aggregation function. Methods based on such flat aggregation usually define $\psi_{aggr}$ that captures simple correlations among nodes. 
For example,~\citep{niepert2016learning} directly concatenates node embeddings together and utilize standard convolutional neural networks as aggregation function to generate graph representation.~\citep{dai2016discriminative} employs a simple element-wise summation operation to define $\psi_{aggr}$, and learns graph embedding by summing all embeddings of individual nodes. 

In addition, some methods apply recurrent neural networks (RNNs) for representing graphs. Some typical methods first sample a number of graph sequences from the input network, and then apply RNN-based autoencoders to generate embedding for each graph sequence. The final graph representation is obtained by either averaging~\citep{jin2018discriminative} or concatenating~\citep{taheri2018learning} these graph sequence embeddings.

\subsection{Hierarchical Aggregation}
In contrast to flat aggregation, the motivation behind \textit{hierarchical} aggregation is to preserve the hierarchical structure that might be presented in the subgraph by aggregating neighborhood information via a hierarchical way. 
~\citep{bruna2013spectral} and~\citep{defferrard2016convolutional} attempt to utilize such hierarchical structure of networks by combining convolutional neural networks with graph coarsening. The main idea behind them is to stack multiple graph coarsening and convolutional layers. In each layer, they first apply graph cluster algorithms to group nodes, and then merge node embeddings within each cluster using element-wise max-pooling. After clustering, they generate a new coarse network by stacking embeddings of clusters together, which is again fed into convolutional layers and the same process repeats. Clusters in each layer can be viewed as subgraphs, and cluster algorithms are used to learn the assignment matrix of subgraphs, so that the hierarchical structure of network is also propagated through layers. Although these methods work well in certain applications, they actually follow a two-stage fashion, where the stages of clustering and embedding may not reinforce each other. 

To avoid this limitation, DiffPool~\citep{ying2018hierarchical} proposes an end-to-end model that does not depend on a deterministic clustering subroutine. The layer-wise propagation rule is formulated as below:
\begin{equation}
\mathbf{M}^{(k+1)}={\mathbf{C}^{(k)}}^T\mathbf{Z}^{(k)}, \ \ \ 
\mathbf{A}^{(k+1)}={\mathbf{C}^{(k)}}^T\mathbf{A}^{(k)}\mathbf{C}^{(k)},
\label{eq12}
\end{equation}
where $\mathbf{Z}^{(k)}\in \mathbb{R}^{N_{k}\times{D}}$ denotes node embeddings, $\mathbf{C}^{(k)}\in \mathbb{R}^{N_{N_k \times{N_{k+1}}}}$ is the cluster assignment matrix learned from the previous layer. The goal of the left equation is to generate the $(k+1)$-th coarser network embedding $\mathbf{M}^{(k+1)}$ by aggregating node embeddings according to cluster assignment $\mathbf{C}^{(k)}$; while the right equation is to learn a new coarsened adjacency matrix $\mathbf{A}^{(k+1)}\in \mathbb{R}^{N_{k+1}\times{N_{k+1}}}$ from the previous adjacency matrix $\mathbf{A}^{(k)}$, which stores the similarity between each pair of clusters. Here, instead of applying deterministic clustering algorithm to learn $\mathbf{C}^{(k)}$, they adopt graph neural networks (GNNs) to learn it. Specifically, they use two separate GNNs on the input embedding matrix $\mathbf{M}^{(k)}$ and coarsened adjacency matrix $\mathbf{A}^{(k)}$  to generate assignment matrix $\mathbf{C}^{(k)}$ and embedding matrix $\mathbf{Z}^{(k)}$, respectively. Formally, $\mathbf{Z}^{(k)}=\textsc{GNN}_{k,embed}(\mathbf{A}^{(k)}, \mathbf{M}^{(k)})$, and $\mathbf{C}^{(k)}=softmax(\textsc{GNN}_{k,pool}(\mathbf{A}^{(k)}, \mathbf{M}^{(k)}))$. The two steps could reinforce each other to improve the performance. DiffPool may suffer from computational issues brought by the computation of soft clustering assignment, which is further addressed in~\citep{cangea2018towards}.

\section{Applications}
The representations learned from networks can be easily applied to downstream machine learning models for further analysis on social networks. Some common applications include node classification, link prediction, anomaly detection and clustering.

\subsection{Node Classification}
In social networks, people are often associated with semantic labels with respect to certain aspects of them, such as affiliations, interests or beliefs. However, in real-world scenarios, people are usually partially or sparsely labeled, since labeling is expensive and time consuming. The goal of node classification is to predict labels of unlabeled nodes in networks by leveraging their connections with the labeled ones considering the network structure. 
According to~\citep{bhagat2011node}, existing methods can be classified into two categories, e.g., random walk based, and feature extraction based methods. The former aims to propagate labels with random walks~\citep{baluja2008video}, while the latter targets to extract features from a node's surrounding information and network statistics.

In general, network representation approach follows the second principle. A number of existing network representation models, like~\citep{yang2015network,wang2016structural,liao2018attributed}, focus on extracting node features from network using representation learning techniques, and then apply machine learning classifiers like support vector machine, naive bayes classifiers, and logistic regression for prediction. In contrast to separating the steps of node embedding and node classification, some recent work~\citep{hamilton2017inductive,dai2016discriminative,monti2017geometric} designs a end-to-end framework to combine the two tasks, so that the discriminative information inferred from labels can directly benefit the learning of network embedding. 

\subsection{Link Prediction}
Social networks are not necessarily complete as some links might be missing. For example, friendship links between two users in a social network can be missing even they actually know each other in real world. The goal of link prediction is to infer the existence of new interactions or emerging links between users in the future, based on the observed links and the network evolution mechanism~\citep{lu2011link,al2011survey,liben2007link}. In network embedding, an effective model is expected to preserve both network structure and inherent dynamics of the network in the low-dimensional space. In general, the majority of previous work focus on predicting missing links between users under homogeneous network settings~\citep{grover2016node2vec, ou2016asymmetric,zhou2017scalable}, and some efforts also attempt to predict missing links in heterogeneous networks~\citep{liu2017semantic,liu2018distance}. Although beyond the scope of this survey, applying network embedding for building recommender systems~\citep{ying2018graph} may also be a direction that is worth exploring.

\subsection{Anomaly Detection}
Another challenging task in social network analysis is anomaly detection. Malicious activities in social networks, such as spamming, fraud and phishing, can be interpreted as rare or unexpected behaviors that deviate from the majority of normal users. While numerous algorithms have been proposed for spotting anomalies and outliers in networks~\citep{Savage2014Anomaly,Akoglu2015Graph,Liu2017Accelerated}, anomaly detection methods based on network embedding techniques are receiving increasing attentions recently~\citep{Hu2016An,Peng2018ANOMALOUS,Liang2018SEANO}. The discrete and structural information in networks are merged and projected into the continuous latent space, which facilitates the application of various statistical or geometrical algorithms in measuring the degree of isolation or outlierness of network components. In addition, in contrast to detect malicious activities in a static way,~\citep{Sricharan2014Localizing} and~\citep{Yu2018NetWalk} also attempt to study the problem in dynamic networks.

\subsection{Node Clustering}
In addition to the above applications, node clustering is another important network analysis problem. The target of node clustering is to partition a network into a set of clusters (or subgraphs), so that nodes in the same cluster are more similar to each other than those from other clusters. In social networks, such clusters are widely spread in terms of communities, such as groups of people that belong to similar affiliations or have similar interests. Most previous work focuses on clustering networks with various metrics of proximity or connection strength between nodes.  For examples,~\citep{shi2000normalized} and~\citep{ding2001min} seek to maximize the number of connections within clusters while minimize the connections between clusters. Recently, many efforts have resort to network representation techniques for node clustering. Some methods treat embedding and clustering as disjoint tasks, where they first embed nodes to low-dimensional vectors, and then apply traditional clustering algorithms to produce clusters~\citep{tian2014learning,cao2015grarep,wang2017community}. Other methods such as~\citep{tang2016capped} and~\citep{wei2017cross} consider the optimization problem of clustering and network embedding in a unified objective function and generate cluster-induced node embeddings.    

\section{Conclusion and Future Directions}
Recent years have witnessed a surge in leveraging representation learning techniques for network analysis.  
In this survey, we have provided a overview of the recent efforts on this topic. Specifically, we summarize existing techniques into three subgroups based on the type of the core learning modules: representation look-up tables, autoencoders and graph convolutional networks. Although many techniques have been developed for a wide spectrum of social networks analysis problems in the past few years, we believe there still remains many promising directions worth of further exploration.   

\noindent\textbf{Dynamic networks}. Social networks are inherently highly dynamic in real-life scenarios. The overall set of nodes, the underlying network structure, as well as attribute information, might evolve over time. As an example, these elements in real world social networks such as Facebook could correspond to users, connections and personal profiles.  
This property makes existing static learning techniques fail to work properly. Although several methods have been proposed to tackle dynamic networks, they often rely on certain assumptions, such as assuming that the node set is fixed and only deal with dynamics caused by edge deletion and addition~\citep{Li2017Attributed}. Also, the changes in attribute information are rarely considered in existing works.
Therefore, how to design effective and efficient network embedding techniques for truly dynamic networks is still an open question.

\noindent\textbf{Hierarchical network structure}. Most of the existing techniques mainly focus on designing advanced encoding or decoding functions trying to capture node pairwise relationships. Nevertheless, pairwise relations can only provide insights about local neighborhoods, and might not infer global hierarchical network structures, which however is crucial for more complex networks~\citep{Benson2016Higher}. How to design effective network embedding methods that are capable of preserving hierarchical structures of networks is an promising direction for further work.

\noindent\textbf{Heterogeneous networks}. Existing network embedding methods mainly deal with homogeneous networks. However, many relational systems in real-life scenarios can be abstracted as heterogeneous networks with multiple types of nodes or edges. In this case, it is hard to evaluate semantic proximity between different network elements in the low-dimensional space. While some work have investigated the use of metapaths~\citep{dong2017metapath2vec,Huang2017Heterogeneous} to approximate semantic similarity for heterogeneous network embedding, many tasks on heterogeneous networks have not been fully evaluated. Learning embeddings for heterogeneous networks is still at the early stage, more comprehensive techniques are needed to fully capture the relations between different types of network elements, towards modeling more complex real systems.

\noindent\textbf{Scalability}. Although deep learning based network embedding methods have achieved substantial performances due to their great capacities, they still suffer from the problem of efficiency. This problem will become more severe when dealing with real-life massive datasets with billions of nodes and edges. Designing deep representation learning frameworks that are scalable for real network datasets is another driving factor to advance the research on this domain. In addition, similar to using GPUs for traditional deep models built on grid structured data, developing computational paradigms for large-scale network processing could be an alternative way towards efficiency improvement~\citep{bronstein2017geometric}. 

\noindent\textbf{Interpretability}. Despite the superior performances achieved by deep models, one fundamental limitation of them is the lack of interpretability~\citep{liu2018interpretation}. Different dimensions in the embedding space usually have no specific meaning, thus it is difficulty to comprehend the underlying factors that have been preserved in the latent space. Since the interpretability aspect of machine learning models is receiving more and more attentions recently~\citep{montavon2018methods,du2018techniques}, it might also be important to explore how to understand the representation learning outcome, how to develop interpretable network representation learning models, as well as how to utilize interpretation to improve the representation models. Answering these questions is helpful for learning more meaningful and task-specific embeddings towards various social network analysis problems. 

\bibliographystyle{frontiersinSCNS_ENG_HUMS} 

\begin{thebibliography}{85}
\providecommand{\natexlab}[1]{#1}
\expandafter\ifx\csname urlstyle\endcsname\relax
  \providecommand{\doi}[1]{doi:\discretionary{}{}{}#1}\else
  \providecommand{\doi}{doi:\discretionary{}{}{}\begingroup
  \urlstyle{rm}\Url}\fi
\providecommand{\selectlanguage}[1]{\relax}
\providecommand{\bibAnnoteFile}[1]{%
  \IfFileExists{#1}{\begin{quotation}\noindent\textsc{Key:} #1\\
  \textsc{Annotation:}\ \input{#1}\end{quotation}}{}}
\providecommand{\bibAnnote}[2]{%
  \begin{quotation}\noindent\textsc{Key:} #1\\
  \textsc{Annotation:}\ #2\end{quotation}}

\bibitem[{Akoglu et~al.(2015)Akoglu, Tong, and Koutra}]{Akoglu2015Graph}
Akoglu, L., Tong, H., and Koutra, D. (2015).
\newblock Graph based anomaly detection and description: a survey.
\newblock \emph{Data Mining and Knowledge Discovery} 29, 626--688
\bibAnnoteFile{Akoglu2015Graph}

\bibitem[{Al~Hasan and Zaki(2011)}]{al2011survey}
Al~Hasan, M. and Zaki, M.~J. (2011).
\newblock A survey of link prediction in social networks.
\newblock In \emph{Social Network Data Analytics} (Springer). 243--275
\bibAnnoteFile{al2011survey}

\bibitem[{Baluja et~al.(2008)Baluja, Seth, Sivakumar, Jing, Yagnik, Kumar
  et~al.}]{baluja2008video}
Baluja, S., Seth, R., Sivakumar, D., Jing, Y., Yagnik, J., Kumar, S., et~al.
  (2008).
\newblock Video suggestion and discovery for youtube: taking random walks
  through the view graph.
\newblock In \emph{International Conference on World Wide Web} (ACM), 895--904
\bibAnnoteFile{baluja2008video}

\bibitem[{Belkin and Niyogi(2002)}]{belkin2002laplacian}
Belkin, M. and Niyogi, P. (2002).
\newblock Laplacian eigenmaps and spectral techniques for embedding and
  clustering.
\newblock In \emph{Advances in Neural Information Processing Systems}. 585--591
\bibAnnoteFile{belkin2002laplacian}

\bibitem[{Bengio et~al.(2013)Bengio, Courville, and
  Vincent}]{bengio2013representation}
Bengio, Y., Courville, A., and Vincent, P. (2013).
\newblock Representation learning: A review and new perspectives.
\newblock \emph{IEEE Transactions on Pattern Analysis and Machine Intelligence}
  35, 1798--1828
\bibAnnoteFile{bengio2013representation}

\bibitem[{Bengio et~al.(2009)}]{bengio2009learning}
Bengio, Y. et~al. (2009).
\newblock Learning deep architectures for ai.
\newblock \emph{Foundations and trends{\textregistered} in Machine Learning} 2,
  1--127
\bibAnnoteFile{bengio2009learning}

\bibitem[{Benson et~al.(2016)Benson, Gleich, and Leskovec}]{Benson2016Higher}
Benson, A.~R., Gleich, D.~F., and Leskovec, J. (2016).
\newblock Higher-order organization of complex networks.
\newblock \emph{Science} 353, 163
\bibAnnoteFile{Benson2016Higher}

\bibitem[{Bhagat et~al.(2011)Bhagat, Cormode, and
  Muthukrishnan}]{bhagat2011node}
Bhagat, S., Cormode, G., and Muthukrishnan, S. (2011).
\newblock Node classification in social networks.
\newblock In \emph{Social Network Data Analytics} (Springer). 115--148
\bibAnnoteFile{bhagat2011node}

\bibitem[{Bronstein et~al.(2017)Bronstein, Bruna, LeCun, Szlam, and
  Vandergheynst}]{bronstein2017geometric}
Bronstein, M.~M., Bruna, J., LeCun, Y., Szlam, A., and Vandergheynst, P.
  (2017).
\newblock Geometric deep learning: going beyond euclidean data.
\newblock \emph{IEEE Signal Processing Magazine} 34, 18--42
\bibAnnoteFile{bronstein2017geometric}

\bibitem[{Bruna et~al.(2013)Bruna, Zaremba, Szlam, and
  LeCun}]{bruna2013spectral}
Bruna, J., Zaremba, W., Szlam, A., and LeCun, Y. (2013).
\newblock Spectral networks and locally connected networks on graphs.
\newblock \emph{arXiv preprint arXiv:1312.6203}
\bibAnnoteFile{bruna2013spectral}

\bibitem[{Bullinaria and Levy(2007)}]{bullinaria2007extracting}
Bullinaria, J.~A. and Levy, J.~P. (2007).
\newblock Extracting semantic representations from word co-occurrence
  statistics: A computational study.
\newblock \emph{Behavior research methods} 39, 510--526
\bibAnnoteFile{bullinaria2007extracting}

\bibitem[{Cangea et~al.(2018)Cangea, Veli{\v{c}}kovi{\'c}, Jovanovi{\'c}, Kipf,
  and Li{\`o}}]{cangea2018towards}
Cangea, C., Veli{\v{c}}kovi{\'c}, P., Jovanovi{\'c}, N., Kipf, T., and Li{\`o},
  P. (2018).
\newblock Towards sparse hierarchical graph classifiers.
\newblock \emph{arXiv preprint arXiv:1811.01287}
\bibAnnoteFile{cangea2018towards}

\bibitem[{Cao et~al.(2015)Cao, Lu, and Xu}]{cao2015grarep}
Cao, S., Lu, W., and Xu, Q. (2015).
\newblock Grarep: Learning graph representations with global structural
  information.
\newblock In \emph{ACM International Conference on Information and Knowledge
  Management} (ACM), 891--900
\bibAnnoteFile{cao2015grarep}

\bibitem[{Cao et~al.(2016)Cao, Lu, and Xu}]{cao2016deep}
Cao, S., Lu, W., and Xu, Q. (2016).
\newblock Deep neural networks for learning graph representations.
\newblock In \emph{AAAI Conference on Artificial Intelligence} (AAAI),
  1145--1152
\bibAnnoteFile{cao2016deep}

\bibitem[{Chang et~al.(2015)Chang, Han, Tang, Qi, Aggarwal, and
  Huang}]{chang2015heterogeneous}
Chang, S., Han, W., Tang, J., Qi, G.-J., Aggarwal, C.~C., and Huang, T.~S.
  (2015).
\newblock Heterogeneous network embedding via deep architectures.
\newblock In \emph{ACM SIGKDD International Conference on Knowledge Discovery
  and Data Mining} (ACM), 119--128
\bibAnnoteFile{chang2015heterogeneous}

\bibitem[{Chen et~al.(2010)Chen, Wang, and Wang}]{chen2010scalable}
Chen, W., Wang, C., and Wang, Y. (2010).
\newblock Scalable influence maximization for prevalent viral marketing in
  large-scale social networks.
\newblock In \emph{ACM SIGKDD International Conference on Knowledge Discovery
  and Data Mining} (ACM), 1029--1038
\bibAnnoteFile{chen2010scalable}

\bibitem[{Dai et~al.(2016)Dai, Dai, and Song}]{dai2016discriminative}
Dai, H., Dai, B., and Song, L. (2016).
\newblock Discriminative embeddings of latent variable models for structured
  data.
\newblock In \emph{International Conference on Machine Learning}. 2702--2711
\bibAnnoteFile{dai2016discriminative}

\bibitem[{Defferrard et~al.(2016)Defferrard, Bresson, and
  Vandergheynst}]{defferrard2016convolutional}
Defferrard, M., Bresson, X., and Vandergheynst, P. (2016).
\newblock Convolutional neural networks on graphs with fast localized spectral
  filtering.
\newblock In \emph{Advances in Neural Information Processing Systems}.
  3844--3852
\bibAnnoteFile{defferrard2016convolutional}

\bibitem[{Dietz et~al.(2007)Dietz, Bickel, and
  Scheffer}]{dietz2007unsupervised}
Dietz, L., Bickel, S., and Scheffer, T. (2007).
\newblock Unsupervised prediction of citation influences.
\newblock In \emph{International Conference on Machine learning} (ACM),
  233--240
\bibAnnoteFile{dietz2007unsupervised}

\bibitem[{Ding et~al.(2001)Ding, He, Zha, Gu, and Simon}]{ding2001min}
Ding, C.~H., He, X., Zha, H., Gu, M., and Simon, H.~D. (2001).
\newblock A min-max cut algorithm for graph partitioning and data clustering.
\newblock In \emph{IEEE International Conference on Data Mining} (IEEE),
  107--114
\bibAnnoteFile{ding2001min}

\bibitem[{Doersch(2016)}]{doersch2016tutorial}
Doersch, C. (2016).
\newblock Tutorial on variational autoencoders.
\newblock \emph{arXiv preprint arXiv:1606.05908}
\bibAnnoteFile{doersch2016tutorial}

\bibitem[{Dong et~al.(2017)Dong, Chawla, and Swami}]{dong2017metapath2vec}
Dong, Y., Chawla, N.~V., and Swami, A. (2017).
\newblock metapath2vec: Scalable representation learning for heterogeneous
  networks.
\newblock In \emph{ACM SIGKDD International Conference on Knowledge Discovery
  and Data Mining} (ACM), 135--144
\bibAnnoteFile{dong2017metapath2vec}

\bibitem[{Du et~al.(2018{\natexlab{a}})Du, Wang, Song, Lu, and
  Wang}]{du2018dynamic}
Du, L., Wang, Y., Song, G., Lu, Z., and Wang, J. (2018{\natexlab{a}}).
\newblock Dynamic network embedding: An extended approach for skip-gram based
  network embedding.
\newblock In \emph{International Joint Conference on Artificial Intelligence}.
  2086--2092
\bibAnnoteFile{du2018dynamic}

\bibitem[{Du et~al.(2018{\natexlab{b}})Du, Liu, and Hu}]{du2018techniques}
Du, M., Liu, N., and Hu, X. (2018{\natexlab{b}}).
\newblock Techniques for interpretable machine learning.
\newblock \emph{arXiv preprint arXiv:1808.00033}
\bibAnnoteFile{du2018techniques}

\bibitem[{Gao and Huang(2018)}]{gao2018deep}
Gao, H. and Huang, H. (2018).
\newblock Deep attributed network embedding.
\newblock In \emph{International Joint Conference on Artificial Intelligence}.
  3364--3370
\bibAnnoteFile{gao2018deep}

\bibitem[{Grover and Leskovec(2016)}]{grover2016node2vec}
Grover, A. and Leskovec, J. (2016).
\newblock node2vec: Scalable feature learning for networks.
\newblock In \emph{ACM SIGKDD International Conference on Knowledge Discovery
  and Data Mining} (ACM), 855--864
\bibAnnoteFile{grover2016node2vec}

\bibitem[{Guo et~al.(2018)Guo, Xu, and Chen}]{guo2018spine}
Guo, J., Xu, L., and Chen, E. (2018).
\newblock Spine: Structural identity preserved inductive network embedding.
\newblock \emph{arXiv preprint arXiv:1802.03984}
\bibAnnoteFile{guo2018spine}

\bibitem[{Guo et~al.(2014)Guo, Zhang, Zhu, Chi, and Gong}]{guo2014two}
Guo, Z., Zhang, Z.~M., Zhu, S., Chi, Y., and Gong, Y. (2014).
\newblock A two-level topic model towards knowledge discovery from citation
  networks.
\newblock \emph{IEEE Transactions on Knowledge and Data Engineering} 26,
  780--794
\bibAnnoteFile{guo2014two}

\bibitem[{Hamilton et~al.(2017{\natexlab{a}})Hamilton, Ying, and
  Leskovec}]{hamilton2017inductive}
Hamilton, W., Ying, Z., and Leskovec, J. (2017{\natexlab{a}}).
\newblock Inductive representation learning on large graphs.
\newblock In \emph{Advances in Neural Information Processing Systems}.
  1024--1034
\bibAnnoteFile{hamilton2017inductive}

\bibitem[{Hamilton et~al.(2017{\natexlab{b}})Hamilton, Ying, and
  Leskovec}]{hamilton2017representation}
Hamilton, W.~L., Ying, R., and Leskovec, J. (2017{\natexlab{b}}).
\newblock Representation learning on graphs: Methods and applications.
\newblock \emph{arXiv preprint arXiv:1709.05584}
\bibAnnoteFile{hamilton2017representation}

\bibitem[{Haussler(1999)}]{haussler1999convolution}
Haussler, D. (1999).
\newblock \emph{Convolution kernels on discrete structures}.
\newblock Tech. rep., Technical report, Department of Computer Science,
  University of California at Santa Cruz
\bibAnnoteFile{haussler1999convolution}

\bibitem[{Henaff et~al.(2015)Henaff, Bruna, and LeCun}]{henaff2015deep}
Henaff, M., Bruna, J., and LeCun, Y. (2015).
\newblock Deep convolutional networks on graph-structured data.
\newblock \emph{arXiv preprint arXiv:1506.05163}
\bibAnnoteFile{henaff2015deep}

\bibitem[{Hinton and Salakhutdinov(2006)}]{hinton2006reducing}
Hinton, G.~E. and Salakhutdinov, R.~R. (2006).
\newblock Reducing the dimensionality of data with neural networks.
\newblock \emph{Science} 313, 504--507
\bibAnnoteFile{hinton2006reducing}

\bibitem[{Hu et~al.(2016)Hu, Aggarwal, Ma, and Huai}]{Hu2016An}
Hu, R., Aggarwal, C.~C., Ma, S., and Huai, J. (2016).
\newblock An embedding approach to anomaly detection.
\newblock In \emph{IEEE International Conference on Data Engineering}. 385--396
\bibAnnoteFile{Hu2016An}

\bibitem[{Huang et~al.(2019)Huang, Song, Yang, and Hu}]{huang2019large}
Huang, X., Song, Q., Yang, F., and Hu, X. (2019).
\newblock Large-scale heterogeneous feature embedding.
\newblock In \emph{AAAI Conference on Artificial Intelligence}
\bibAnnoteFile{huang2019large}

\bibitem[{Huang and Mamoulis(2017)}]{Huang2017Heterogeneous}
Huang, Z. and Mamoulis, N. (2017).
\newblock Heterogeneous information network embedding for meta path based
  proximity.
\newblock \emph{arXiv preprint arXiv:1701.05291}
\bibAnnoteFile{Huang2017Heterogeneous}

\bibitem[{Jin et~al.(2018)Jin, Song, and Hu}]{jin2018discriminative}
Jin, H., Song, Q., and Hu, X. (2018).
\newblock Discriminative graph autoencoder.
\newblock In \emph{IEEE International Conference on Big Knowledge (ICBK)}
\bibAnnoteFile{jin2018discriminative}

\bibitem[{Kipf and Welling(2016)}]{kipf2016variational}
Kipf, T.~N. and Welling, M. (2016).
\newblock Variational graph auto-encoders.
\newblock \emph{arXiv preprint arXiv:1611.07308}
\bibAnnoteFile{kipf2016variational}

\bibitem[{Le and Mikolov(2014)}]{le2014distributed}
Le, Q. and Mikolov, T. (2014).
\newblock Distributed representations of sentences and documents.
\newblock In \emph{International Conference on Machine Learning}. 1188--1196
\bibAnnoteFile{le2014distributed}

\bibitem[{Leskovec et~al.(2007)Leskovec, Adamic, and
  Huberman}]{leskovec2007dynamics}
Leskovec, J., Adamic, L.~A., and Huberman, B.~A. (2007).
\newblock The dynamics of viral marketing.
\newblock \emph{ACM Transactions on the Web} 1, 5
\bibAnnoteFile{leskovec2007dynamics}

\bibitem[{Levy and Goldberg(2014)}]{levy2014neural}
Levy, O. and Goldberg, Y. (2014).
\newblock Neural word embedding as implicit matrix factorization.
\newblock In \emph{Advances in Neural Information Processing Systems}.
  2177--2185
\bibAnnoteFile{levy2014neural}

\bibitem[{Li et~al.(2017{\natexlab{a}})Li, Wang, Yang, and
  Odagaki}]{li2017variation}
Li, H., Wang, H., Yang, Z., and Odagaki, M. (2017{\natexlab{a}}).
\newblock Variation autoencoder based network representation learning for
  classification.
\newblock In \emph{Proceedings of ACL 2017, Student Research Workshop}
\bibAnnoteFile{li2017variation}

\bibitem[{Li et~al.(2017{\natexlab{b}})Li, Dani, Hu, Tang, Chang, and
  Liu}]{Li2017Attributed}
Li, J., Dani, H., Hu, X., Tang, J., Chang, Y., and Liu, H.
  (2017{\natexlab{b}}).
\newblock Attributed network embedding for learning in a dynamic environment.
\newblock In \emph{ACM Conference on Information and Knowledge Management}.
  387--396
\bibAnnoteFile{Li2017Attributed}

\bibitem[{Li et~al.(2015)Li, Zhang, and Tan}]{li2015real}
Li, Y., Zhang, D., and Tan, K.-L. (2015).
\newblock Real-time targeted influence maximization for online advertisements.
\newblock \emph{Proceedings of the VLDB Endowment} 8, 1070--1081
\bibAnnoteFile{li2015real}

\bibitem[{Liang et~al.(2018)Liang, Jacobs, Sun, and
  Parthasarathy}]{Liang2018SEANO}
Liang, J., Jacobs, P., Sun, J., and Parthasarathy, S. (2018).
\newblock Semi-supervised embedding in attributed networks with outliers.
\newblock In \emph{SIAM International Conference on Data Mining} (SIAM),
  153--161
\bibAnnoteFile{Liang2018SEANO}

\bibitem[{Liao et~al.(2018)Liao, He, Zhang, and Chua}]{liao2018attributed}
Liao, L., He, X., Zhang, H., and Chua, T.-S. (2018).
\newblock Attributed social network embedding.
\newblock \emph{IEEE Transactions on Knowledge and Data Engineering}
\bibAnnoteFile{liao2018attributed}

\bibitem[{Liben-Nowell and Kleinberg(2007)}]{liben2007link}
Liben-Nowell, D. and Kleinberg, J. (2007).
\newblock The link-prediction problem for social networks.
\newblock \emph{Journal of the American Society for Information Science and
  Technology} 58, 1019--1031
\bibAnnoteFile{liben2007link}

\bibitem[{Liu et~al.(2017{\natexlab{a}})Liu, Huang, and
  Hu}]{Liu2017Accelerated}
Liu, N., Huang, X., and Hu, X. (2017{\natexlab{a}}).
\newblock Accelerated local anomaly detection via resolving attributed
  networks.
\newblock In \emph{International Joint Conference on Artificial Intelligence}.
  2337--2343
\bibAnnoteFile{Liu2017Accelerated}

\bibitem[{Liu et~al.(2018{\natexlab{a}})Liu, Huang, Li, and
  Hu}]{liu2018interpretation}
Liu, N., Huang, X., Li, J., and Hu, X. (2018{\natexlab{a}}).
\newblock On interpretation of network embedding via taxonomy induction.
\newblock In \emph{ACM SIGKDD International Conference on Knowledge Discovery
  and Data Mining} (KDD)
\bibAnnoteFile{liu2018interpretation}

\bibitem[{Liu et~al.(2017{\natexlab{b}})Liu, Zheng, Zhao, Zhu, Chang, Wu
  et~al.}]{liu2017semantic}
Liu, Z., Zheng, V.~W., Zhao, Z., Zhu, F., Chang, K. C.-C., Wu, M., et~al.
  (2017{\natexlab{b}}).
\newblock Semantic proximity search on heterogeneous graph by proximity
  embedding.
\newblock In \emph{AAAI Conference on Artificial Intelligence}. 154--160
\bibAnnoteFile{liu2017semantic}

\bibitem[{Liu et~al.(2018{\natexlab{b}})Liu, Zheng, Zhao, Zhu, Chang, Wu
  et~al.}]{liu2018distance}
Liu, Z., Zheng, V.~W., Zhao, Z., Zhu, F., Chang, K. C.-C., Wu, M., et~al.
  (2018{\natexlab{b}}).
\newblock Distance-aware dag embedding for proximity search on heterogeneous
  graphs.
\newblock In \emph{AAAI Conference on Artificial Intelligence}
\bibAnnoteFile{liu2018distance}

\bibitem[{L{\"u} and Zhou(2011)}]{lu2011link}
L{\"u}, L. and Zhou, T. (2011).
\newblock Link prediction in complex networks: A survey.
\newblock \emph{Physica A: Statistical Mechanics and Its Applications} 390,
  1150--1170
\bibAnnoteFile{lu2011link}

\bibitem[{Meng et~al.(2019)Meng, Liang, Bao, and Zhang}]{meng2019co}
Meng, Z., Liang, S., Bao, H., and Zhang, X. (2019).
\newblock Co-embedding attributed networks.
\newblock In \emph{ACM International Conference on Web Search and Data Mining}.
  393--401
\bibAnnoteFile{meng2019co}

\bibitem[{Mikolov et~al.(2013)Mikolov, Chen, Corrado, and
  Dean}]{mikolov2013efficient}
Mikolov, T., Chen, K., Corrado, G., and Dean, J. (2013).
\newblock Efficient estimation of word representations in vector space.
\newblock \emph{arXiv preprint arXiv:1301.3781}
\bibAnnoteFile{mikolov2013efficient}

\bibitem[{Montavon et~al.(2018)Montavon, Samek, and
  M{\"u}ller}]{montavon2018methods}
Montavon, G., Samek, W., and M{\"u}ller, K.-R. (2018).
\newblock Methods for interpreting and understanding deep neural networks.
\newblock \emph{Digital Signal Processing}
\bibAnnoteFile{montavon2018methods}

\bibitem[{Monti et~al.(2017)Monti, Boscaini, Masci, Rodola, Svoboda, and
  Bronstein}]{monti2017geometric}
Monti, F., Boscaini, D., Masci, J., Rodola, E., Svoboda, J., and Bronstein,
  M.~M. (2017).
\newblock Geometric deep learning on graphs and manifolds using mixture model
  cnns.
\newblock In \emph{The IEEE Conference on Computer Vision and Pattern
  Recognition}. vol.~1, 3
\bibAnnoteFile{monti2017geometric}

\bibitem[{Natarajan and Dhillon(2014)}]{natarajan2014inductive}
Natarajan, N. and Dhillon, I.~S. (2014).
\newblock Inductive matrix completion for predicting gene--disease
  associations.
\newblock \emph{Bioinformatics} 30, i60--i68
\bibAnnoteFile{natarajan2014inductive}

\bibitem[{Niepert et~al.(2016)Niepert, Ahmed, and
  Kutzkov}]{niepert2016learning}
Niepert, M., Ahmed, M., and Kutzkov, K. (2016).
\newblock Learning convolutional neural networks for graphs.
\newblock In \emph{International Conference on Machine Learning}. 2014--2023
\bibAnnoteFile{niepert2016learning}

\bibitem[{Ou et~al.(2016)Ou, Cui, Pei, Zhang, and Zhu}]{ou2016asymmetric}
Ou, M., Cui, P., Pei, J., Zhang, Z., and Zhu, W. (2016).
\newblock Asymmetric transitivity preserving graph embedding.
\newblock In \emph{ACM SIGKDD International Conference on Knowledge Discovery
  and Data Mining} (ACM), 1105--1114
\bibAnnoteFile{ou2016asymmetric}

\bibitem[{Peng et~al.(2017)Peng, Wang, and Xie}]{peng2017social}
Peng, S., Wang, G., and Xie, D. (2017).
\newblock Social influence analysis in social networking big data:
  Opportunities and challenges.
\newblock \emph{IEEE Network} 31, 11--17
\bibAnnoteFile{peng2017social}

\bibitem[{Peng et~al.(2018)Peng, Luo, Li, Liu, and Zheng}]{Peng2018ANOMALOUS}
Peng, Z., Luo, M., Li, J., Liu, H., and Zheng, Q. (2018).
\newblock Anomalous: A joint modeling approach for anomaly detection on
  attributed networks.
\newblock In \emph{International Joint Conference on Artificial Intelligence}.
  3513--3519
\bibAnnoteFile{Peng2018ANOMALOUS}

\bibitem[{Perozzi et~al.(2014)Perozzi, Al-Rfou, and
  Skiena}]{perozzi2014deepwalk}
Perozzi, B., Al-Rfou, R., and Skiena, S. (2014).
\newblock Deepwalk: Online learning of social representations.
\newblock In \emph{ACM SIGKDD International Conference on Knowledge Discovery
  and Data Mining} (ACM), 701--710
\bibAnnoteFile{perozzi2014deepwalk}

\bibitem[{Qiu et~al.(2018)Qiu, Dong, Ma, Li, Wang, and Tang}]{qiu2018network}
Qiu, J., Dong, Y., Ma, H., Li, J., Wang, K., and Tang, J. (2018).
\newblock Network embedding as matrix factorization: Unifying deepwalk, line,
  pte, and node2vec.
\newblock In \emph{ACM International Conference on Web Search and Data Mining}.
  459--467
\bibAnnoteFile{qiu2018network}

\bibitem[{Savage et~al.(2014)Savage, Zhang, Yu, Chou, and
  Wang}]{Savage2014Anomaly}
Savage, D., Zhang, X., Yu, X., Chou, P., and Wang, Q. (2014).
\newblock Anomaly detection in online social networks.
\newblock \emph{Social Networks} 39, 62--70
\bibAnnoteFile{Savage2014Anomaly}

\bibitem[{Shi et~al.(2019)Shi, Hu, Zhao, and Philip}]{shi2019heterogeneous}
Shi, C., Hu, B., Zhao, W.~X., and Philip, S.~Y. (2019).
\newblock Heterogeneous information network embedding for recommendation.
\newblock \emph{IEEE Transactions on Knowledge and Data Engineering} 31(2),
  357--370
\bibAnnoteFile{shi2019heterogeneous}

\bibitem[{Shi and Malik(2000)}]{shi2000normalized}
Shi, J. and Malik, J. (2000).
\newblock Normalized cuts and image segmentation.
\newblock \emph{IEEE Transactions on Pattern Analysis and Machine Intelligence}
  22, 888--905
\bibAnnoteFile{shi2000normalized}

\bibitem[{Song et~al.(2006)Song, Tseng, Lin, and Sun}]{song2006personalized}
Song, X., Tseng, B.~L., Lin, C.-Y., and Sun, M.-T. (2006).
\newblock Personalized recommendation driven by information flow.
\newblock In \emph{International ACM SIGIR Conference on Research and
  Development in Information Retrieval} (ACM), 509--516
\bibAnnoteFile{song2006personalized}

\bibitem[{Sricharan and Das(2014)}]{Sricharan2014Localizing}
Sricharan, K. and Das, K. (2014).
\newblock Localizing anomalous changes in time-evolving graphs.
\newblock In \emph{ACM SIGMOD International Conference on Management of Data}
  (ACM), 1347--1358
\bibAnnoteFile{Sricharan2014Localizing}

\bibitem[{Taheri et~al.(2018)Taheri, Gimpel, and
  Berger-Wolf}]{taheri2018learning}
Taheri, A., Gimpel, K., and Berger-Wolf, T. (2018).
\newblock Learning graph representations with recurrent neural network
  autoencoders.
\newblock \emph{KDD Deep Learning Day}
\bibAnnoteFile{taheri2018learning}

\bibitem[{Tang et~al.(2015)Tang, Qu, Wang, Zhang, Yan, and Mei}]{tang2015line}
Tang, J., Qu, M., Wang, M., Zhang, M., Yan, J., and Mei, Q. (2015).
\newblock Line: Large-scale information network embedding.
\newblock In \emph{International Conference on World Wide Web}. 1067--1077
\bibAnnoteFile{tang2015line}

\bibitem[{Tang et~al.(2016)Tang, Nie, and Jain}]{tang2016capped}
Tang, M., Nie, F., and Jain, R. (2016).
\newblock Capped lp-norm graph embedding for photo clustering.
\newblock In \emph{ACM Multimedia Conference} (ACM), 431--435
\bibAnnoteFile{tang2016capped}

\bibitem[{Tang and Yang(2012)}]{tang2012ranking}
Tang, X. and Yang, C.~C. (2012).
\newblock Ranking user influence in healthcare social media.
\newblock \emph{ACM Transactions on Intelligent Systems and Technology} 3, 73
\bibAnnoteFile{tang2012ranking}

\bibitem[{Thekumparampil et~al.(2018)Thekumparampil, Wang, Oh, and
  Li}]{thekumparampil2018attention}
Thekumparampil, K.~K., Wang, C., Oh, S., and Li, L.-J. (2018).
\newblock Attention-based graph neural network for semi-supervised learning.
\newblock \emph{arXiv preprint arXiv:1803.03735}
\bibAnnoteFile{thekumparampil2018attention}

\bibitem[{Tian et~al.(2014)Tian, Gao, Cui, Chen, and Liu}]{tian2014learning}
Tian, F., Gao, B., Cui, Q., Chen, E., and Liu, T.-Y. (2014).
\newblock Learning deep representations for graph clustering.
\newblock In \emph{AAAI Conference on Artificial Intelligence}. 1293--1299
\bibAnnoteFile{tian2014learning}

\bibitem[{Velickovic et~al.(2017)Velickovic, Cucurull, Casanova, Romero, Lio,
  and Bengio}]{velickovic2017graph}
Velickovic, P., Cucurull, G., Casanova, A., Romero, A., Lio, P., and Bengio, Y.
  (2017).
\newblock Graph attention networks.
\newblock \emph{arXiv preprint arXiv:1710.10903}
\bibAnnoteFile{velickovic2017graph}

\bibitem[{Vishwanathan et~al.(2010)Vishwanathan, Schraudolph, Kondor, and
  Borgwardt}]{vishwanathan2010graph}
Vishwanathan, S. V.~N., Schraudolph, N.~N., Kondor, R., and Borgwardt, K.~M.
  (2010).
\newblock Graph kernels.
\newblock \emph{Journal of Machine Learning Research} 11, 1201--1242
\bibAnnoteFile{vishwanathan2010graph}

\bibitem[{Wang et~al.(2016)Wang, Cui, and Zhu}]{wang2016structural}
Wang, D., Cui, P., and Zhu, W. (2016).
\newblock Structural deep network embedding.
\newblock In \emph{ACM SIGKDD International Conference on Knowledge Discovery
  and Data Mining} (ACM), 1225--1234
\bibAnnoteFile{wang2016structural}

\bibitem[{Wang et~al.(2017)Wang, Cui, Wang, Pei, Zhu, and
  Yang}]{wang2017community}
Wang, X., Cui, P., Wang, J., Pei, J., Zhu, W., and Yang, S. (2017).
\newblock Community preserving network embedding.
\newblock In \emph{AAAI Conference on Artificial Intelligence}. 203--209
\bibAnnoteFile{wang2017community}

\bibitem[{Wei et~al.(2017)Wei, Xu, Cao, and Yu}]{wei2017cross}
Wei, X., Xu, L., Cao, B., and Yu, P.~S. (2017).
\newblock Cross view link prediction by learning noise-resilient representation
  consensus.
\newblock In \emph{International Conference on World Wide Web}. 1611--1619
\bibAnnoteFile{wei2017cross}

\bibitem[{Yang et~al.(2015)Yang, Liu, Zhao, Sun, and Chang}]{yang2015network}
Yang, C., Liu, Z., Zhao, D., Sun, M., and Chang, E.~Y. (2015).
\newblock Network representation learning with rich text information.
\newblock In \emph{International Joint Conference on Artificial Intelligence}.
  2111--2117
\bibAnnoteFile{yang2015network}

\bibitem[{Yang et~al.(2016)Yang, Cohen, and Salakhutdinov}]{yang2016revisiting}
Yang, Z., Cohen, W., and Salakhutdinov, R. (2016).
\newblock Revisiting semi-supervised learning with graph embeddings.
\newblock In \emph{International Conference on Machine Learning}. 40--48
\bibAnnoteFile{yang2016revisiting}

\bibitem[{Ying et~al.(2018{\natexlab{a}})Ying, He, Chen, Eksombatchai,
  Hamilton, and Leskovec}]{ying2018graph}
Ying, R., He, R., Chen, K., Eksombatchai, P., Hamilton, W.~L., and Leskovec, J.
  (2018{\natexlab{a}}).
\newblock Graph convolutional neural networks for web-scale recommender
  systems.
\newblock \emph{arXiv preprint arXiv:1806.01973}
\bibAnnoteFile{ying2018graph}

\bibitem[{Ying et~al.(2018{\natexlab{b}})Ying, You, Morris, Ren, Hamilton, and
  Leskovec}]{ying2018hierarchical}
Ying, R., You, J., Morris, C., Ren, X., Hamilton, W.~L., and Leskovec, J.
  (2018{\natexlab{b}}).
\newblock Hierarchical graph representation learning withdifferentiable
  pooling.
\newblock \emph{arXiv preprint arXiv:1806.08804}
\bibAnnoteFile{ying2018hierarchical}

\bibitem[{Yu et~al.(2018)Yu, Cheng, Aggarwal, Zhang, Chen, and
  Wang}]{Yu2018NetWalk}
Yu, W., Cheng, W., Aggarwal, C.~C., Zhang, K., Chen, H., and Wang, W. (2018).
\newblock Netwalk: A flexible deep embedding approach for anomaly detection in
  dynamic networks.
\newblock In \emph{ACM SIGKDD International Conference on Knowledge Discovery
  and Data Mining}. 2672--2681
\bibAnnoteFile{Yu2018NetWalk}

\bibitem[{Zhou et~al.(2017)Zhou, Liu, Liu, Liu, and Gao}]{zhou2017scalable}
Zhou, C., Liu, Y., Liu, X., Liu, Z., and Gao, J. (2017).
\newblock Scalable graph embedding for asymmetric proximity.
\newblock In \emph{AAAI Conference on Artificial Intelligence}. 2942--2948
\bibAnnoteFile{zhou2017scalable}

\end{thebibliography}



\end{document}